# PERFORMANCE MEASUREMENT OF CLOUD COMPUTING SERVICES

Sinung Suakanto[1], Suhono H Supangkat[2], Suhardi[3] and Roberd Saragih[4]

[1,2,3]School of Electrical Engineering and Informatics, Bandung Institute of Technology, Bandung, Indonesia

[1]mr.sinung.suakanto@gmail.com
[2]suhono@stei.itb.ac.id,
[3]suhardi@stei.itb.ac.id

[4]Departement of Mathematic, Bandung Institute of Technology, Bandung, Indonesia

[4]roberd@math.itb.ac.id

**ABSTRACT**

*Cloud computing today has now been growing as new technologies and new business models. In distributed technology perspective, cloud computing most like client-server services like web-based or web-service but it used virtual resources to execute. Currently, cloud computing relies on the use of an elastic virtual machine and the use of network for data exchange. We conduct an experimental setup to measure the quality of service received by cloud computing customers. Experimental setup done by creating a HTTP service that runs in the cloud computing infrastructure. We interest to know about the impact of increasing the number of users on the average quality received by users. The qualities received by user measured within two parameters consist of average response times and the number of requests time out. Experimental results of this study show that increasing the number of users has increased the average response time. Similarly, the number of request time out increasing with increasing number of users. It means that the qualities of service received by user are decreasing also. We found that the impact of the number of users on the quality of service is no longer in linear trend. The results of this study can be used as a reference model for the network operator in performing services in which a certain number of users in order to obtain optimal quality services.*

## 1. INTRODUCTION

Currently, the usage of internet is growth rapidly and most of the internet traffic is generated by the World Wide Web (WWW) and the associated protocol Hypertext Transfer Protocol (HTTP). With the rapid growth of internet has brought the development of new technologies such as grid and cloud computing [10]. This technology would successfully implement if it supported by the growth of broadband access network. Currently, bandwidth of network access has grown rapidly also with various technologies such as 3G, HSDPA, LTE and others. It may also led into the emergence of new network operators including cloud computing provider in infrastructure (IaaS), platform (PaaS) or service (SaaS) [11].

Many problem faces by the internet users are almost concern about perceived quality by users [6],[7]. Provider likely ignores the average quality of service received by user. Most providers focused on pushing the growth number of users but don't pay attention at the declining quality received by the average user. For that, this study measured and analyzes how the impact of increasing number of the user with the average quality received at the user. These results may be used as model reference for the provider to prepare their services much better.

Cloud computing is one that utilizes virtualization technology to represent the presence of a device such as server [13]. The user is only required to create an account with the service provider in order to use and utilize services [12]. The virtual resources as the part of the cloud computing such as virtual machine (VM) will execute either virtual routers or virtualized service elements. These virtual resources will be combined together to create some specific cloud computing services. In terms of application services, applications running in cloud computing is not much different from traditional service on the web server. The difference between them is that the cloud computing service is run on the virtual machine, not in physical machine as usual. It may lead into creating much identical application service that could managed more efficiently.

Some cloud computing infrastructure was deploying as industry / commercial purpose or academic purposes. We can found open source cloud computing services such as Eucalyptus [4] and OpenNebula[1]. Various types of cloud computing architecture have also been developed as public or private cloud. We can see one of it that developed by Intel as seen in Figure 1 [15]. In the picture, we can see that the network used as a connector between all nodes or components involved building a cloud computing infrastructure. The network it self used to exchange information or coordination between nodes. Therefore it needs a high speed network to provide best services.

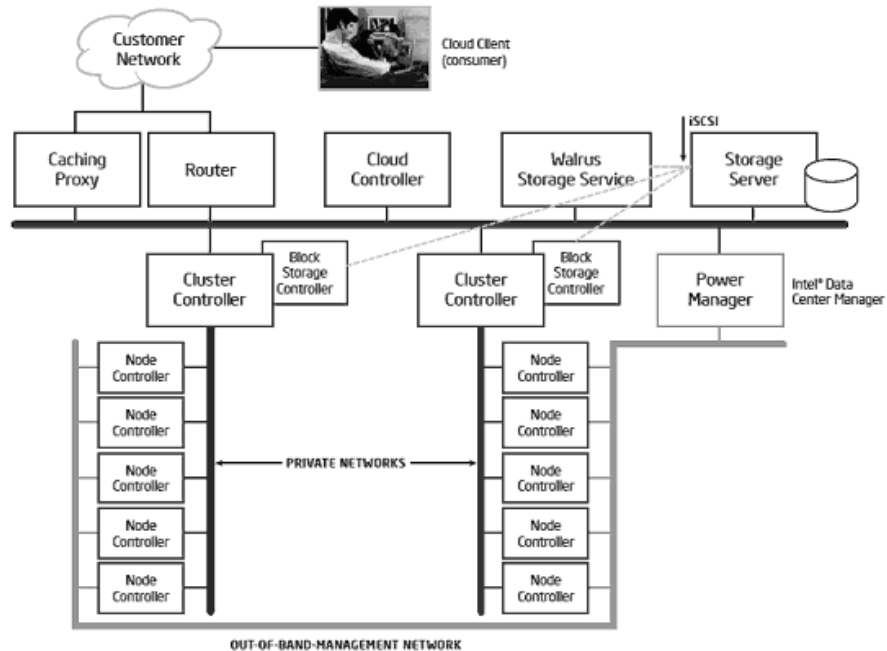

Figure 1. Example of Cloud Computing Infrastructure Deployment

Cloud computing encapsulates a variety of services wherein the software applications running on a cloud infrastructure are offered as services to the consumers [12]. Cloud computing has evolved to run services like web-based application or Web Services [3],[12]. Web-based application or web service is a service that uses HTTP protocol to exchange data between the clients with the server. HTTP protocol has best-effort characteristics that there is no assurance of quality service. Cloud computing itself is expected have quality of service guarantee to meet the SLA (Service Level Agreement) between user and provider [11],[5].

At a previous work, we had been proposed a method for maintaining quality of service for applications such as for cloud computing services. One of parameter being maintained is that the average response time value must be above a certain value [14]. The method was developed using the FTR-HTTP (Finite Time Response-HTTP) which is a method to determine the response time is expected to be above a certain value [14].

Because of HTTP characteristic, they have dynamic performances that can unpredictable result for end-to-end communication. Dynamic changes for the transmission throughput may cause by flow control mechanism. The value of the throughput is dependent on the probability of packet loss and round trip time (RTT) like formula proposed by Padhye [9]. Several studies have been done more focus on the empirical measurement of the web based application [5], [16]. The analysis is done on various HTTP traffic parameters, e.g., inter-session timings, inter-arrival timings, request message sizes, response code, and number of transactions. The reported results can be useful for building synthetic workloads for simulation and benchmarking purposes. Some studies made to measure the performance of HTTP service on WiFi networks [16]. They did estimate the maximum number of users can be supported in a WiFi-based network when the users are consuming basic services [16].

## 2. RESEARCH METHOD

We are interesting with problems about how the impact of increasing the number of users on the average quality received by each user. This study focused on conduct the measurement and analysis of service performance of cloud computing applications. Research methodology could be seen as in the Figure 2. First, we proposed hypothesis that we presume the increasing number of users will decrease the average performance received by users.

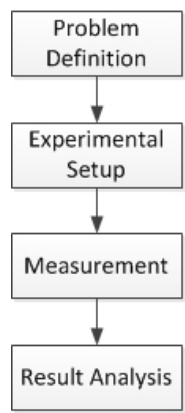

Figure 2. Research Methodology

## 2.1. Experimental Setup

To restrict our discussion, this research proposes to use specific model of cloud computing architecture with specific service running on it. As we mention before, we know that cloud computing architecture may diverse in implementation both for public or private cloud computing. Generally, the users of public cloud computing will spread in different places geographically. In the context of some specific business model, cloud computing users can originate from the same network or region. For example, institution or organization put their data centre in some specific external cloud provider. With this model, most of the users of the service or underlying application came from the same network or region. Most of them are the member of its organization. The concept of using cloud computing services like this can be described as shown in Figure 4. This model is our basic model cloud architecture used for measurement.

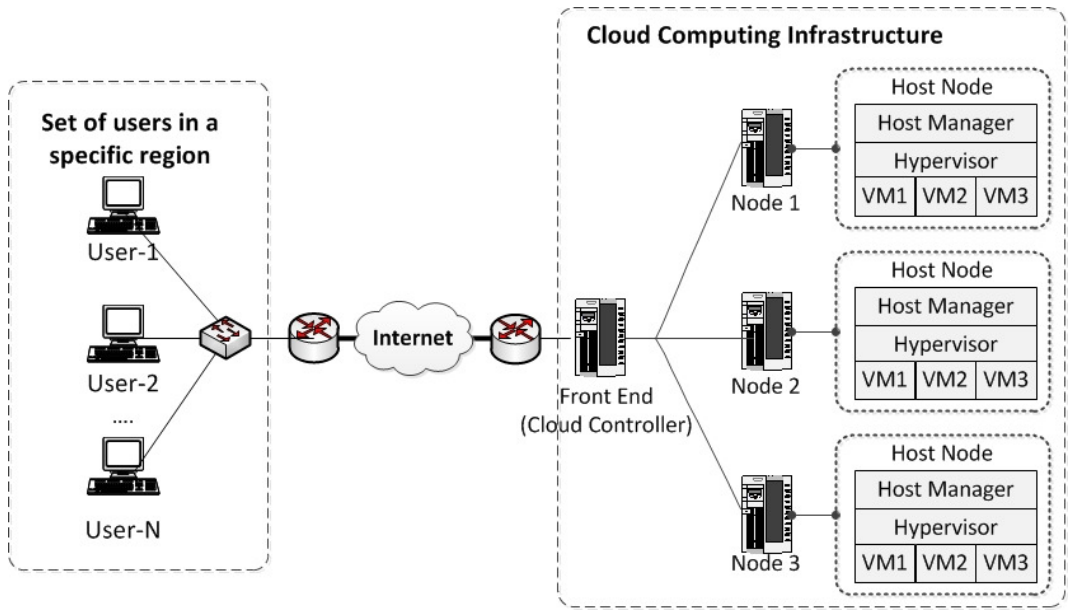

Figure 3. Architecture Model of Public Cloud Computing Usage

Applications that are widely developed in cloud computing today use HTTP protocol like web-application and web service application. In order to accommodate this requirement, our experimental setup will also use the HTTP-based applications that run in the cloud computing architecture. This could be setup as our model implementation that can be seen as in the Figure 5.

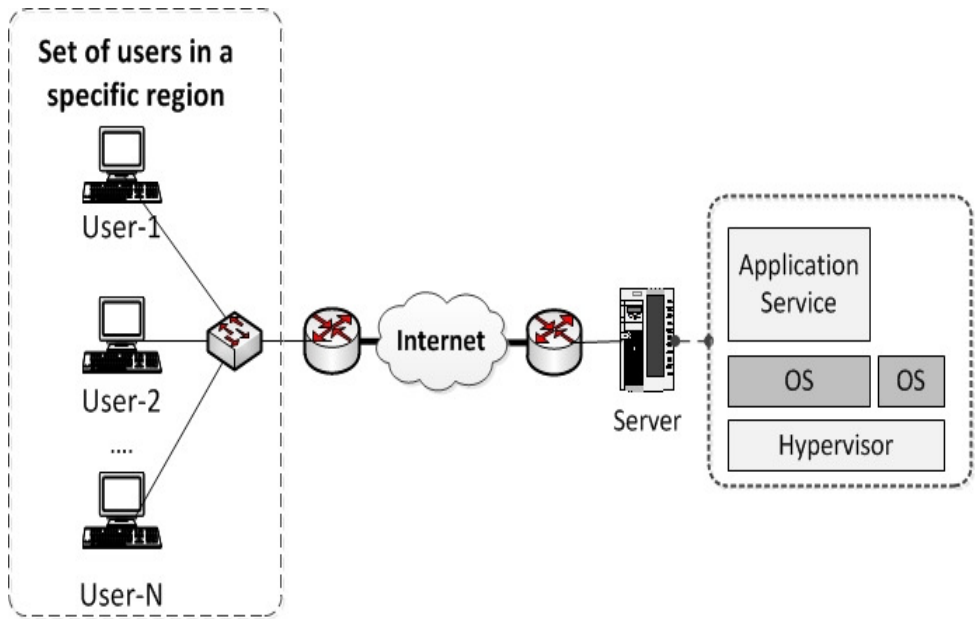

Figure 4. Experimental Setup Model Implementation

In order to measure performance at this model, it seems difficult to make direct measurements of many users who do access simultaneously at the same timeframe. Therefore, to support these scenarios we create our own services at the clients that will emulate many users accessing the service cloud. It seems like to an online traffic generator. For this purpose, we need to know how users work when accessing some specific application running in the server. Also, we need to know how HTTP protocol works in order to help us understate and analyze experimental result discussion.

We must denote if the fundamental working of services that uses HTTP protocol is based on request-response. For each transaction or data exchange identify as single request-response process. When user takes doing something on the application (dialogue-interaction via graphical user interface) then the application will send a request to the server. The server would process it and sent response back to the client. Thus the process will be carried out repeatedly. It is a random processes or stochastic processes.

To describe how users use the service in the cloud computing or similar service, we use a stochastic approach. Request and response are in random. The interval between the requests would not in the same period but randomly in time (stochastic process). Similarly, if there are more users then it will be more requests to be sent to the server. It can be described as shown in Figure 5.

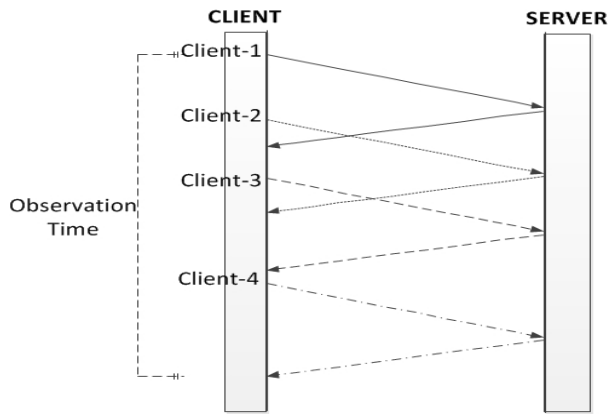

Figure 5. Multiple Request on Specific Observation Time

To realize this concept into emulation setup, system we use two parameters that is:

- Number of users (N)
- Number of requests per period for each user (R). For the interval between request, we use normal distribution.

The assumptions used in this setup simulation include:

- Each user only access into single application service. It means that all traffic directed to the same server cloud computing. None of user access to other servers except into cloud computing server.
- Number of requests per period for each user is assumed equal but the interval between requests is in a random process. For example, each customer has average 10 requests per minute. Interval between first request and next request still in random process.
- All of users conduct transactions continuously during the period of observation. None of user suddenly stops or pause during the period of observation.

The number of users describes the number of active users during the period of observation. This amount does not reflecting the total number of users in the system.

## 2.2. Experimental Scenarios

To provide more comprehensive results, we propose to setup experiments with 4 different network regions for user to access server as described in Table 1.

Table 1. Type of Network Region for Origin Source

| Network Region | Number of Hops | Internet Connection | Type of Network User |
|---|---|---|---|
| 1 | 13 | Wireless Broadband Access | A |
| 2 | 13 | Broadband Campus Network | B |
| 3 | 13 | DSL | A |
| 4 | 3 | Broadband Campus Network | B |

For these network regions, we were interested to see how many hops to know distance between clients into cloud computing server. Naturally, the more hop show greater distance between the clients to the server. The more the number of hops it will show how many network components through which the data packets from client to server that may lead into more networks delay. We are also interested to see the characteristics of network used by the client for internet connection, whether internet access using broadband or others.

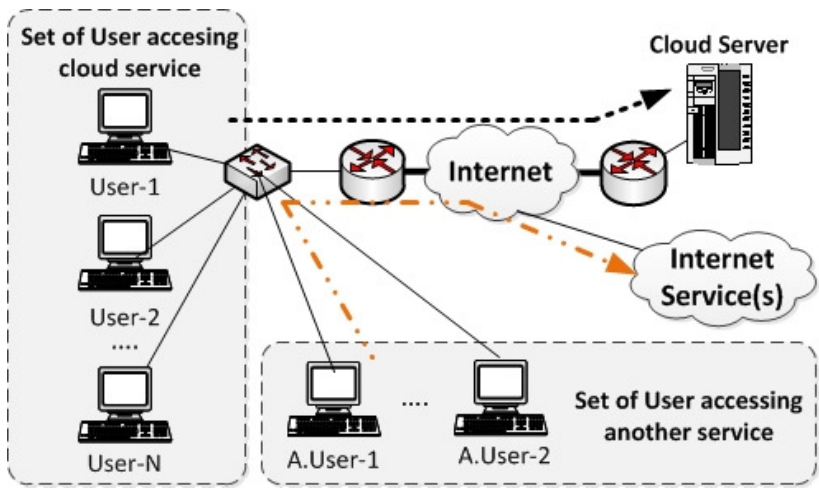

Figure 6. Network Region Type B  Also, we are interested to know if the client use of the internet exclusively or whether they used for shared. The meaning with exclusive here represent that all users or most users will only have access to cloud computing. If many users who are use access exclusively to the cloud server, we categorized with Type A. For this type, we can describe it like shown at Figure 4. Otherwise, if there are some users who were also using the internet to access to a different place then we categorized with Type B. This kind of type could be described like shown at Figure 6.
.
For services running on cloud computing infrastructure, we built some services application that running for search engine services utility. Users would access it as front end application via HTTP protocol like conservative web-based application to search and get result based on keyword. We put it on application server bundled with Web Server that running in specific virtual machine (VM) on cloud computing. The difference with the usual traditional web server architecture is if the web server running on single computer physical called server. Whereas in this experiment, the web server running on a virtual machine running (VM). Every single physical computer denote as node can load more than one virtual machine.

## 3. RESULT AND ANALYSIS

### 3.1. Result

In this section, we will discuss the results of measurements to access services running on the cloud computing infrastructure. For this research purposes, we are interested to know about the impact of increasing number of user to the average performance received by each user. The architecture used in this experiment is as shown in Figure 5. For cloud computing server, we have successfully built an infrastructure cloud computing using Eucalyptus and have run Virtual

Machines (VM). In the Virtual Machine will be installed with application service which will be tested and measured about its performance when accessing from user perspective.

Setup for this experiment are services at client running during certain periods to access services in the cloud computing. To reduce the variation of the data resulted; we use similar request and the similar response. It is done to avoid the variation caused by a different of packet data length. With the same request and response then we can know that the length of data packets both for request and response has same respectively.

We use parameter like number of requests per user (R) is same for each user. For this prior experiment, we use 1 request per minute. The experiments were carried out repeatedly by varying the number of users to get the average performance at the client (quality received by user). For quality parameters measured, we are interested to measure the response time and the ratio of the number of requests that time out. Response time indicates how long the distance between when the user sends a request until accepted the response from the server. If response time is taking a long time then it indicates that the service is not in good performance perceived by user. The results of measurements of changes in the number of users on the average response time that users receive can be seen as in the Figure 7.

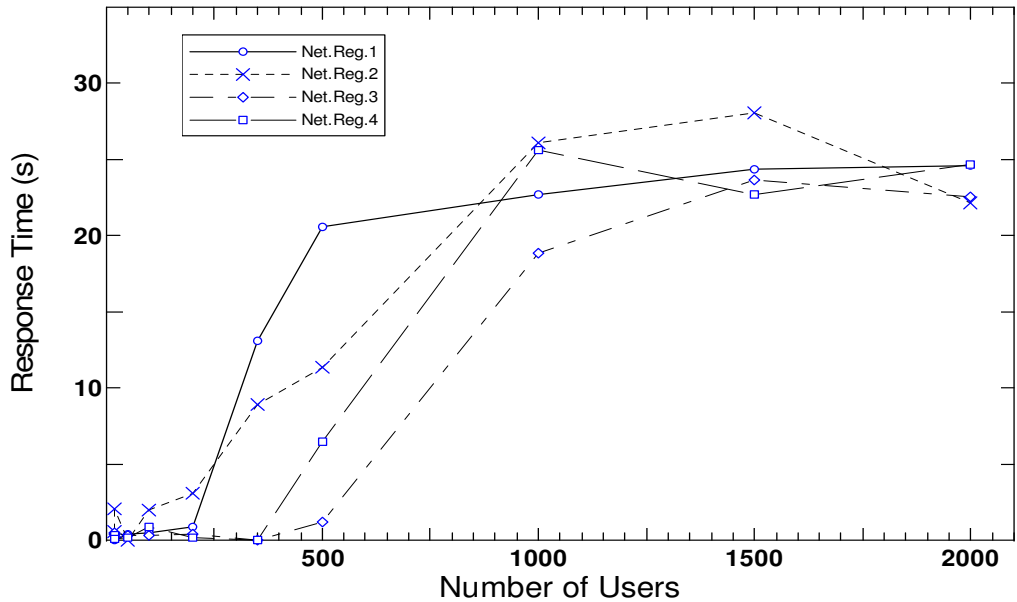

Figure 7. Average Response Time between Changes of Number of Users

We are also interested in looking for the ratio of the number of requests that have time out. By default, we calculate this ratio by the following formula:

$$PTOR = \frac{Number\ of\ request\ time\ out}{Total\ number\ of\ request}.100\,\% \qquad (1)$$

PTOR : Percentage of Time Out Request

16

These ratios express how much requests that experience on time out. Users who experience this time out would not be happy because the transaction failed after such a long wait. Time out is caused by many things such as the server is too busy processed the request because of high queuing request. It may be also caused by the number of requests that is lost in the network because of data flooding in particular network node. As we know that the HTTP protocol has a maximum value of time out. When the time out value was passed then it indicates that there is a request that is too long in the network or on the server. To avoid waiting time for the client then the client can determine the value of its time out. In this experiment, we set time out for 60 seconds. For the measurement results that indicate the number of requests that timeout can be seen as in the Figure 8.

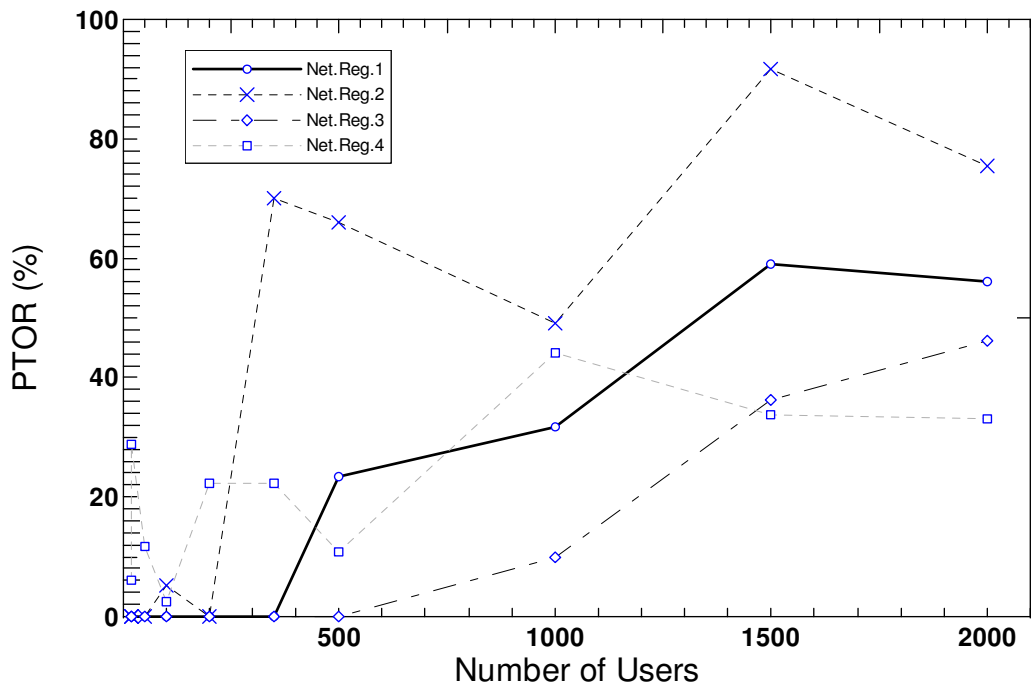

Figure 8. Number of Ratio Request Time Out between Changes of Number of Users

## 3.2. Analysis

The experimental results showed that increasing the number of users will decrease the performance received by the user. In this case shown by the decrease in the average response time receive by the user. It is also demonstrated by the increasing amount of requests that have time out. Response time here does not reflect the response time of processing on the server only but also involves a delay caused by network delay. So the response time obtained from this experiment in the seconds-order rather than in milliseconds-order. We assume that both the network and server have the same contribution to the delay.

$$Total\ response\ time = delay\ processing\ at\ server + delay\ network \quad (1)$$

Intuitively, we know that increasing the number of users will increase the average response time received by users. Based on the above experimental results depicted at the Figure 7 shows us that the trend doesn't have the linear trend. We observe that when the number of users is still small example below 200 users seen that the response time is still in good results. However, when the

17

number of users was in the middle range (for example: between 200 until 500), the average response time increasing significantly. When the number of user is more than 1000 indicates that response time is very high but the growth is not as fast as before (small gradient) or in saturate condition. We observe the slow growth or saturated event caused by compensation from the increasing number of requests that have time out (see Figure 8). Request time out of this experience can be defined mathematically as a response request time over than 60 second (We use HTTP Time out setting for 60 second). If this time out request is taken into account in calculating the average response time, it will produce a graph where the average response time with value closer in the 60 seconds.

For these initial experiments, we analyzed the response time a little produced are more likely to approach the S-shape curve or sigmoid function than the linear form. A sigmoid function can be shown like in the Figure 9. This statement is supported by our initial experimental result that described when first period resulted value in slow growth. Then in the middle period are significantly increased (fast growth), whereas in the last period increased but was no longer significant or denoted as slow growth or saturate condition.

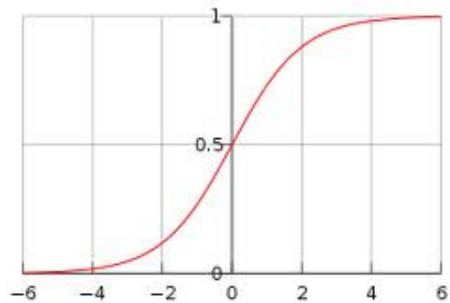

For the graph in the Figure 8, we only see trend that increasing number of users followed with increasing number of requests time out (RTO). But the results were not as smooth-trend as in the graph from Figure 7. From this perspective, we can observer further that the network I and III showed good graphics results with increasing trend. While the Network Region II and IV shows that tend to be random dynamics. We can observe again that this could be related to network type region. Network Region I and III are Type A and result on better tendency. While the network region II and IV (Type B) showed more variation and random dynamics. This is presumably because the network type B is also used for access for other purposes. This fluctuation may cause due to the density of data packets that have various destination.

## 4. CONCLUSIONS

This study has showed us that service measurements in cloud computing based on empirical method. With this scenario, it can be concluded empirically that increasing number of users could increasing average response time received by users. The trend tends to sigmoid-like function. The number of requests that experiencing time out also increases with increasing number of users.
The results of these measurements can be used as recommendations or reference model for cloud provider when deploy their services to the users. How much number of users could be served to provide cloud computing services in accepted quality? The increasing number of user that is not followed by increasing in capacity will lead to deterioration in quality received by users. Provider should be performing these measurements like this before providing services to its users.

**Authors**

**Sinung Suakanto** received the Bachelor of Electrical Engineering, from Bandung Institute of Technology in 2004. He received the Master Degree from Bandung Institute of Technology in 2008. Since 2008, he has been with Bandung Institute of Technology as PhD student. His current int erests are Networking, Cloud Computing and Quality of Service (QoS). Currently he is member of IEEE and IEICE-Japan.

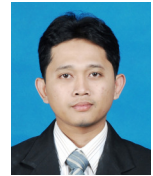

**Suhono H Supangkat** obtained his Ph.D degree from University of Electro-Communications, Tokyo, Japan in 1998, and Bachelor of Engineering degree from Bandung Institute of Technology in 1986. Master of Physical Sciences and Engineering - Electrical Engineering from Meisei University, Japan in 1994. Currently, he is a Professor in Information Technology Research Centre of School of Electrical Engineering and Informatics Engineering at Bandung Institute of Technology, Indonesia. His specialization includes Cloud Computing, Distributed System, and Service Orientation Architecture (SOA).

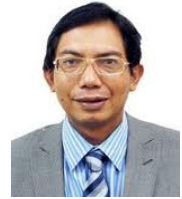

**Suhardi** obtained his Ph.D degree from Technische Universitat Berlin, Berlin, Germany in 1998, Bachelor of Electrical Engineering degree and Master of Electrical Engineering degrees in 1986 and 1993 respectively from Bandung Institute of Technology, Indonesia. Currently, he is a Associate Professor in Information Technology Research Centre of School of Electrical Engineering and Informatics Engineering at Bandung Institute of Technology, Indonesia. His specialization includes Network, Optimization and Cloud Computing.

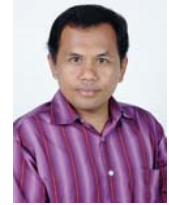

**Roberd Saragih** obtained his Ph.D degree from Keio University, Japan in 1998, Bachelor of Mathematic degree and Master of Instrumentation and Control degrees in 1986 and 1993 respectively from Bandung Institute of Technology, Indonesia. Currently, he is a Professor in Department of Math at Bandung Institute of Technology, Indonesia. His specialization includes Stochastic and Robust Control System, Optimization and Distributed System.

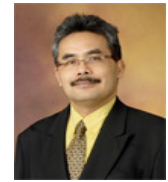